\documentclass[preprint2]{aastex}

\shorttitle{NIR Polarization and Variation of {\it Fermi}/LAT Sources}
\shortauthors{Fujiwara et al.}

\begin{document}


\title{Polarization and Variation of Near-IR Light from {\it Fermi}/LAT $\gamma$-ray Sources}


\author{M. Fujiwara\altaffilmark{1}, Y. Matsuoka\altaffilmark{1}, and N. Ienaka\altaffilmark{2}}


\altaffiltext{1}{Graduate School of Science, Nagoya University, Furo-cho, Chikusa-ku, Nagoya 464-8602, Japan;
matsuoka@a.phys.nagoya-u.ac.jp}
\altaffiltext{2}{Institute of Astronomy, The University of Tokyo, Osawa 2-21-1, Mitaka, Tokyo 181-0015, Japan}


\begin{abstract}
We present the results of our follow-up observation program of $\gamma$-ray sources detected by the Large Area Telescope (LAT) on board 
the {\it Fermi Gamma-ray Space Telescope}.
26 blazars and 39 sources unidentified at other wavelengths were targeted at IRSF 1.4 m telescope equipped with the SIRIUS/SIRPOL imager 
and polarimeter.
$H$-band magnitudes of the blazars at the epoch of 2010 Dec -- 2011 Feb are presented, which reveal clear flux variation
since the Two Micron All Sky Survey observations and can be useful data for variation analyses of these objects in longer periods.
We also find that nearly half of the $\gamma$-ray blazars are highly ($>$10 \%) polarized in near-infrared wavelengths.
Combining the polarization and variation properties, most ($\sim$90 \%) of the blazars are clearly distinguished from all other types 
of objects at high Galactic latitudes.
On the other hand, we find only one highly polarized and/or variable object in the fields of unidentified sources.
This object is a counterpart of the optical variable source PQV1 J131553.00$-$073302.0 and the radio source NVSS J131552$-$073301,
and is a promising candidate of new $\gamma$-ray blazars.
From the measured polarization and variation statistics, we conclude that most of the {\it Fermi}/LAT unidentified sources are not 
likely similar types of objects to the known $\gamma$-ray blazars.
\end{abstract}


\keywords{BL Lacertae objects: general --- gamma rays: galaxies --- galaxies: active --- galaxies: jets --- polarization --- quasars: general}



\section{Introduction}

A new era of blazar studies has arrived with the advent of the {\it Fermi Gamma-ray Space Telescope} ({\it Fermi}).
{\it Fermi} has been carrying out an all-sky survey with its main instrument Large Area Telescope (LAT) since the science mission
phase started in 2008.
The first {\it Fermi}-LAT catalog \citep[1FGL;][]{{2010ApJS..188..405A}} lists 1451 $\gamma$-ray detections, in which 821 sources are associated
(or identified) with objects found at other wavelengths.
The majority of the associated objects are active galactic nuclei (AGNs) dominated by blazars, while other extragalactic sources as
well as Galactic objects such as pulsars and supernova remnants make smaller contributions.
The remaining 630 sources are left unidentified in 1FGL.

Blazars are considered to be AGNs whose jets are aligned with the observer's line-of-sight.
Emission from the relativistically boosted jets dominates observed flux, resulting in two broad peaks in the spectral
energy distribution (SED); one at radio to X-rays arising from synchrotron emission of accelerated, high-energy particles, and another
at X-ray to $\gamma$-ray arising from the inverse Compton scattering of the lower energy photons.
Because of these emission mechanisms, blazars are characterized by strong radio, X-ray, and $\gamma$-ray radiations as well as
high polarization and variation across the entire SED.
Since blazars dominate the $\gamma$-ray sky at high Galactic latitudes, the {\it Fermi} all-sky survey is expected to shed
new light on this relatively rare and poorly understood population.

An investigation of blazar radiation mechanisms (hence the intrinsic SED) is important not only for revealing the nature of
blazars themselves, but also for measuring the extragalactic background light (EBL).
Since high-energy photons from blazars interact with optical to near-infrared (IR) EBL in the intergalactic space, 
observations of distant blazars can be used to infer the EBL spectrum if the intrinsic blazar SED is precisely known.
With this indirect method, \citet{aharonian06} obtained the significantly-lower upper limits of near-IR EBL than
those derived from the direct measurements by, e.g., \citet{matsumoto05}.
Recently \citet{matsuoka11a} succeeded in a direct measurement of optical EBL by re-analyzing the {\it Pioneer 10/11} data,
which is on the smooth extension of the near-IR upper limits obtained by \citet{aharonian06}.

In this paper, we present the results of our near-IR follow-up observation program of $\gamma$-ray blazars and unidentified sources
in 1FGL.
Despite of the wealth of information there, near-IR wavelength of the whole AGN population is still poorly understood
\citep[e.g.,][]{matsuoka07,matsuoka08,matsuoka11b}.
We aim to quantify the most distinct features of blazars in near-IR wavelengths, i.e., polarization and variation,
in part as a benchmark for future observations.
At the same time, we explore the nature of 1FGL unidentified sources by comparing their near-IR properties to the known
$\gamma$-ray blazars.
Dominance of blazars in the $\gamma$-ray sky implies that some of the unidentified sources at high Galactic latitudes are 
similar objects missed in the past surveys at other wavelengths.
Revealing the origin of these unidentified $\gamma$-ray emissions is of greatest importance, therefore many follow-up studies are being
dedicated to this subject \citep[e.g.,][]{ackermann11b}.
We aim to provide complementary information to these follow-up programs from a near-IR view point.

While the second {\it Fermi}-LAT catalog \citep[2FGL;][]{nolan12} has already been released, this paper is based on 1FGL
for consistency with the sample selection of presented observations.
We discuss the new 2FGL identifications of the sample later.
Magnitudes are presented on the Vega-based system throughout this paper.

\section{Observations and Reduction}


The sample consists of blazars and unidentified sources extracted from 1FGL.
We selected targets at high Galactic latitudes ($|b| >$ 20$^{\circ}$), with small position errors ($<$ 5 arcmin with 95-\% confidence
in most cases), and with bright Two Micron All Sky Survey (2MASS\footnote{
This publication makes use of data products from the Two Micron All Sky Survey, which is a joint project of the University of Massachusetts 
and the Infrared Processing and Analysis Center/California Institute of Technology, funded by the National Aeronautics and Space Administration 
and the National Science Foundation.}) magnitudes ($H_{\rm 2MASS} <$ 15 mag) for the blazars whose accurate coordinates are known.
The objects listed in the {\it Fermi}-LAT first AGN catalog \citep[1LAC;][]{abdo10b} were removed from the sample of unidentified sources.
The polarimetric observations were performed with the Infrared Survey Facility (IRSF) 1.4-m telescope at the South Africa Astronomical
Observatory, Sutherland. 
We used the near-IR imaging camera SIRIUS \citep[Simultaneous Infrared Imager for Unbiased Survey;][]{nagashima99,nagayama03} which is
equipped with three 1024 $\times$ 1024 HgCdTe arrays with the pixel scale of 0''.453.
It was designed to obtain 7'.7 $\times$ 7'.7 images in the three bands, $J$ (1.25 $\micron$), $H$ (1.63 $\micron$) and $K_{\rm s}$ 
(2.14 $\micron$), simultaneously.
The instrument provides polarimetric capability called SIRPOL, with an achromatic (1-2.5 $\micron$) wave-plate rotator unit and a polarizer. 
Source fluxes ($F_{000}$, $F_{225}$, $F_{450}$, $F_{675}$) in the four wave-plate angles, 0$^{\circ}$.0, 22$^{\circ}$.5, 45$^{\circ}$.0, 
67$^{\circ}$.5, are measured in consecutive exposures.
The uncertainty of a measured polarization degree is estimated to be less than 0.3\% \citep{2006SPIE.6269E.159K}.

We summarize the observed targets in Table \ref{tab:obs_bzx} (for blazars) and \ref{tab:obs_una} (for unidentified sources).
The observations were carried out in 2010 December and 2011 February. 
Mean total exposure times were 30 min for a blazar and 60 min for an unidentified source, divided into single exposures of 20 sec 
between which the telescope pointing was dithered by 20 arcsec.
Typical seeing during the observations was $\sim$1''.4. 
Dark and twilight-flat images were obtained before and/or after each night of observations.

\begin{table*}
  \caption{Observation journal of blazars}\label{tab:obs_bzx}
  \begin{center}
    \begin{tabular}{llccccc}
      \hline
              & Associated  & Obs. & $H_{\rm 2MASS}$ & $H_{\rm IRSF}$ & $H^{\rm lim}_{\rm IRSF}$       & Polarization\\
      1FGL ID & blazar      & date\tablenotemark{a} &  (mag)          & (mag)          &  (mag)\footnote[2] & (\%)\\
      \hline
      J0021.7$-$2556 & CRATES J0021$-$2550 & Dec 07 & 14.05 $\pm$ 0.03 & 15.06 $\pm$ 0.02 & 15.54   & 13.8 $\pm$ 2.2\\
      J0033.5$-$1921 & RBS 76              & Dec 13 & 14.08 $\pm$ 0.03 & 14.18 $\pm$ 0.01 & 14.86   &  6.8 $\pm$ 1.8\\
      J0038.4$-$2504 & PKS 0035$-$252      & Dec 13 & 13.12 $\pm$ 0.03 & 15.92 $\pm$ 0.07 & 14.84   & ...\\
      J0050.6$-$0928 & PKS 0048$-$09       & Dec 08 & 13.60 $\pm$ 0.03 & 13.17 $\pm$ 0.01 & 15.65   & 15.5 $\pm$ 0.5\\
      J0120.5$-$2700 & PKS 0118$-$272      & Dec 08 & 13.40 $\pm$ 0.03 & 13.38 $\pm$ 0.01 & 15.57   &  6.5 $\pm$ 0.6\\
      J0132.6$-$1655 & PKS 0130$-$17       & Dec 08 & 14.26 $\pm$ 0.05 & 14.90 $\pm$ 0.01 & 15.44   &  6.4 $\pm$ 1.9\\
      J0209.3$-$5229 & BZB J0209$-$5229    & Dec 09 & 13.80 $\pm$ 0.04 & 14.38 $\pm$ 0.01 & 15.95   &  3.6 $\pm$ 1.0\\
      J0210.6$-$5101 & PKS 0208$-$512      & Dec 07 & 12.86 $\pm$ 0.02 & 14.64 $\pm$ 0.01 & 15.66   & 11.6 $\pm$ 1.4\\
      J0238.6$-$3117 & BZB J0238$-$3116    & Dec 09 & 14.43 $\pm$ 0.06 & 14.36 $\pm$ 0.01 & 16.01   &  3.5 $\pm$ 0.9\\
      J0303.5$-$2406 & PKS 0301$-$243      & Dec 07 & 13.69 $\pm$ 0.04 & 13.17 $\pm$ 0.01 & 15.51   &  3.6 $\pm$ 0.5\\
      J0325.9$-$1649 & RBS 421             & Dec 08  & 14.46 $\pm$ 0.04 & 14.78 $\pm$ 0.01 & 15.41   &  3.2 $\pm$ 1.8\\
      J0334.4$-$3727 & CRATES J0334$-$3725 & Dec 07 & 14.00 $\pm$ 0.03 & 13.36 $\pm$ 0.01 & 15.35   &  4.3 $\pm$ 0.7\\
      J0423.2$-$0118 & PKS 0420$-$01       & Dec 07 & 14.53 $\pm$ 0.05 & 15.48 $\pm$ 0.02 & 15.57   &  8.8 $\pm$ 3.0\\
      J0449.5$-$4350 & PKS 0447$-$439      & Dec 07 & 13.20 $\pm$ 0.03 & 11.87 $\pm$ 0.01 & 15.23   &  4.5 $\pm$ 0.3\\
      J0455.6$-$4618 & PKS 0454$-$46       & Dec 11 & 14.82 $\pm$ 0.05 & 15.62 $\pm$ 0.03 & 15.43   & 11.7 $\pm$ 3.9\\
      J0522.8$-$3632 & PKS 0521$-$36       & Dec 12 & 12.21 $\pm$ 0.04 & 12.02 $\pm$ 0.01 & 14.95   & 10.9 $\pm$ 0.4\\
      J0538.8$-$4404 & PKS 0537$-$441      & Dec 12 & 12.38 $\pm$ 0.03 & 11.72 $\pm$ 0.01 & 15.24   & 12.8 $\pm$ 0.3\\
      J0953.0$-$0838 & CRATES J0953$-$0840 & Dec 09 & 14.18 $\pm$ 0.04 & 14.39 $\pm$ 0.01 & 15.35   & $<$ 3.8\\
      J1022.8$-$0115 & BZB J1022$-$0113    & Dec 12 & 14.92 $\pm$ 0.07 & 15.41 $\pm$ 0.02 & 15.89   & $<$ 3.7\\
      J1059.3$-$1132 & PKS B1056$-$113     & Dec 23 & 14.80 $\pm$ 0.05 & 13.85 $\pm$ 0.01 & 15.57   & 11.6 $\pm$ 0.8\\
      J1126.8$-$1854 & PKS 1124$-$186      & Dec 23 & 13.30 $\pm$ 0.03 & 13.57 $\pm$ 0.01 & 15.28   & 12.6 $\pm$ 0.8\\
      J1204.3$-$0714 & CRATES J1204$-$0710 & Feb 14 & 14.12 $\pm$ 0.09 & 13.85 $\pm$ 0.01 & 16.12   &  3.0 $\pm$ 0.7\\
      J2158.8$-$3013 & PKS 2155$-$304      & Dec 13 & 10.76 $\pm$ 0.03 & 10.46 $\pm$ 0.01 & 14.04   &  3.7 $\pm$ 0.3\\
      J2222.5$-$5218 & BZB J2221$-$5225    & Dec 13 & 14.92 $\pm$ 0.09 & 14.40 $\pm$ 0.01 & 15.09   &  9.2 $\pm$ 1.8\\
      J2235.7$-$4817 & PKS 2232$-$488      & Dec 08 & 12.95 $\pm$ 0.03 & 15.47 $\pm$ 0.03 & 15.43   & $<$ 5.8\\
      J2359.0$-$3035 & 1H 2351$-$315       & Dec 13 & 14.72 $\pm$ 0.07 & 14.33 $\pm$ 0.01 & 15.23   &  1.8 $\pm$ 1.5\\
      \hline
    \end{tabular}
    \tablenotetext{a}{The observations were carried out from 2010 December to 2011 February.}
    \tablenotetext{b}{Limiting magnitudes below which photometry error in a single wave-plate image is less than 0.05 mag.}
  \end{center}
\end{table*}

\begin{table}
  \caption{Observation journal of unidentified sources}\label{tab:obs_una}
  \begin{center}
    \begin{tabular}{lcc}
      \hline
              & Obs.                   & $H^{\rm lim}_{\rm IRSF}$ \\
       1FGL ID & date\tablenotemark{a} & (mag)\tablenotemark{b} \\
      \hline
      J0001.9$-$4158  & Dec 26 &         15.58    \\
      J0028.9$-$7028  & Dec 27 &         15.12   \\
      J0032.7$-$5519  & Dec 25 &         15.67    \\
      J0101.0$-$6423  & Dec 27 &         15.68   \\
      J0136.3$-$2220  & Dec 13 &         15.29   \\
      J0143.9$-$5845  & Dec 23, Feb 07 & 15.95    \\
      J0223.0$-$1118  & Feb 13 &         16.16   \\
      J0247.4$-$6003  & Dec 09, Feb 06 & 16.21   \\
      J0311.3$-$0922  & Dec 24 &         15.67   \\
      J0316.3$-$6438  & Dec 23, Feb 11 & 15.96   \\
      J0335.5$-$4501  & Dec 26 &         16.16    \\
      J0345.2$-$2355  & Dec 24 &         15.86    \\
      J0404.5$-$0850  & Dec 22 &         15.44   \\
      J0409.9$-$0357  & Dec 25, Feb 11 & 15.84    \\
      J0439.6$-$0538  & Dec 24 &         15.71   \\
      J0439.8$-$1857  & Dec 24, Feb 06 & 15.84    \\
      J0515.6$-$4404  & Dec 22 &         15.68    \\      
      J0523.5$-$2529  & Feb 07 &         16.67    \\
      J0614.1$-$3328  & Feb 07 &         16.61     \\
      J0828.9$+$0901  & Feb 13 &         16.34     \\
      J1101.3$+$1009  & Feb 07 &         16.73    \\
      J1119.9$-$2205  & Feb 06 &         16.45    \\
      J1124.4$-$3654  & Feb 14 &         16.12      \\
      J1141.8$-$1403  & Feb 11 &         16.65   \\
      J1223.4$-$3034  & Feb 13 &         16.72      \\
      J1231.1$-$1410  & Feb 06 &         16.59    \\
      J1311.7$-$3429  & Feb 14 &         16.15   \\
      J1312.6$+$0048  & Feb 11 &         16.39     \\
      J1315.6$-$0729  & Feb 11 &         16.75     \\
      J1351.8$-$1523  & Feb 06 &         16.71    \\
      J1511.8$-$0513  & Feb 06 &         16.67     \\
      J2118.3$-$3237  & Dec 25 &         15.20      \\
      J2152.4$-$7532  & Dec 25 &         15.53    \\
      J2227.4$-$7804  & Dec 28 &         15.33     \\
      J2228.5$-$1633  & Dec 29 &         14.93    \\
      J2241.9$-$5236  & Dec 25 &         15.39     \\
      J2251.2$-$4928  & Dec 29 &         14.97   \\
      J2330.3$-$4745  & Dec 28 &         15.58      \\
      J2355.9$-$6613  & Dec 28 &         16.07     \\
      \hline
    \end{tabular}
    \tablenotetext{a}{The observations were carried out from 2010 December to \\ 2011 February.}
    \tablenotetext{b}{\ Limiting magnitudes below which photometry error \\ in a single wave-plate image is less than 0.05 mag.}
  \end{center}
\end{table}

\begin{figure*}
\epsscale{2.0}
\plotone{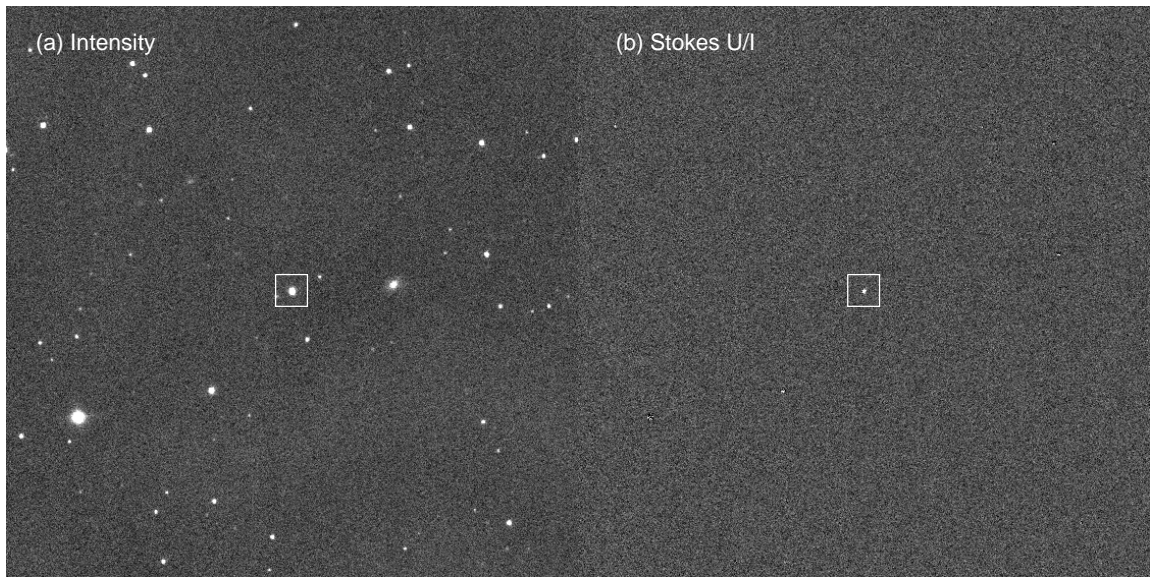}
\caption{IRSF/SIRPOL $H$-band image of intensity (panel a) and Stokes U/I (b) of a field around the blazar
  1FGL J0538.8$-$4404. The blazar is marked with the boxes. \label{fig1}}
\end{figure*}

Data reduction was performed in a standard manner with a dedicated package SIRPOL of the Image Reduction and Analysis 
Facility (IRAF\footnote{
IRAF is distributed by the National Optical Astronomy Observatory, which is operated by the Association of Universities for Research 
in Astronomy (AURA) under cooperative agreement with the National Science Foundation.}), 
including dark subtraction, flat fielding, and substituting bad pixels.
Photometric calibration was achieved by referring to several 2MASS sources within each observed field.
While simultaneous $J$, $H$, and $K_{\rm s}$ band images were obtained, we use $H$-band images in this study because of their
highest quality (signal-to-noise ratios).
We show an example of a reduced $H$-band image in the left panel of Figure \ref{fig1}.

We used the {\it Source Extractor} \citep{1996A&AS..117..393B}, version 2.5, for aperture photometry of detected sources.
Aperture sizes were determined as twice the mean full widths at half maximum of stellar profiles in each image.
We required photometry errors to be less than 0.05 mag for polarization measurements, which defines our limiting magnitudes
$H_{\rm IRSF}^{\rm lim}$ listed in Table \ref{tab:obs_bzx} and \ref{tab:obs_una}.
The average values are $H_{\rm IRSF}^{\rm lim}$  = 15.4 and 16.0 mag for the fields of blazars and unidentified sources, respectively.
Polarization degree $P_{\rm raw}$ was then calculated from measured fluxes $F_{000}$, $F_{225}$, $F_{450}$, $F_{675}$ as follows:
\begin{eqnarray*}
  Q = F_{000} - F_{450} ,\\
  U = F_{225} - F_{675} ,\\
  I = (F_{000} + F_{450} + F_{225} + F_{675})/2 ,\\
  P_{\rm raw} = 100 \times \frac{\sqrt{Q^{2} + U^{2}}}{I} .
\end{eqnarray*}
Here $Q$ and $U$ are the Stokes parameters and $I$ is total intensity. 
A source with strong polarization stands out in Q/I or U/I images as demonstrated in the right panel of Figure \ref{fig1}.

Since $F_{000}$, $F_{225}$, $F_{450}$, $F_{675}$ are not observed simultaneously, temporal variation of the atmospheric condition 
causes additional polarimetry errors to those from the standard sky background noise.
Hence we estimated polarization error $\Delta P$ in a given field from the median $P_{\rm raw}$ value of all the detected sources, 
assuming that most of those sources at high Galactic latitudes have no intrinsic polarization.
Then {\it true} polarization degrees were derived by de-biasing measured $P_{\rm raw}$ \citep{wardle74} following
\begin{eqnarray*}
  P = \sqrt{P_{\rm raw}^{2} - \Delta P^{2}} .
\end{eqnarray*}  
Thus the presented values of polarization and its error should be regarded as the lower and upper limits, respectively, considering
the above assumption on no intrinsic polarization for most of the detected sources.
We provide upper limits of polarization for objects with $P_{\rm raw} < \Delta P$ or $P < \Delta P$.

\section{Results and Discussion}

\begin{figure}
\epsscale{1.0}
\plotone{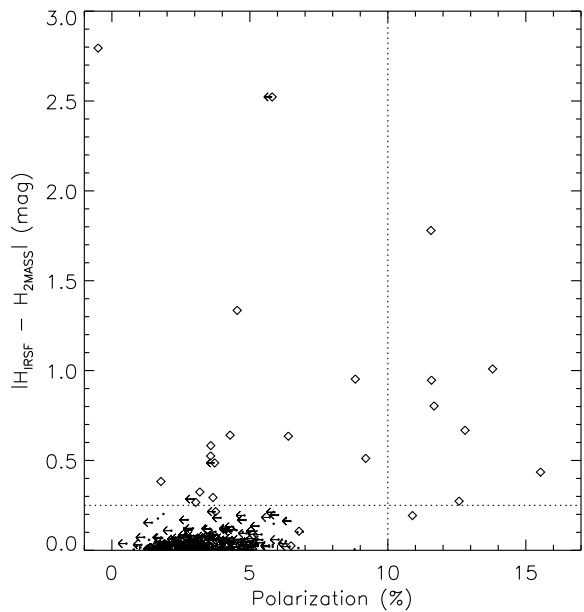}
\caption{Polarization and variation of 326 sources detected in the blazar fields. 
  The diamonds represent the blazars associated to 1FGL sources while the dots represent all other objects. 
  The arrows denote upper limits of polarization. \label{var_stat_bzx}}
\end{figure}

We present the results of photometry and polarimetry measurements for the blazars in Table \ref{tab:obs_bzx}.
Their polarization and variation since 2MASS observations, as well as those for all other sources detected in the blazar fields, 
are plotted in Figure \ref{var_stat_bzx}.
Strong variability of the blazars is evident; 22 out of 26 blazars (85 \%) have $| H_{\rm IRSF} - H_{\rm 2MASS} | > 0.25$ mag
while only one of the other detected sources shows such variation ($\sim$0.3 mag; it is likely a contaminating normal star).
The blazars are also characterized by pronounced polarization; 10 out of 25 blazars (40 \%;
polarization was not measured for a blazar associated to J0038.4$-$2504 due to a large ($>$ 0.05 mag) photometry error) 
have $P >$ 10 \%, while all other types of objects are much less polarized.
The presented fraction, 40 \%, represents a useful benchmark for planning future optical/near-IR polarization follow-up 
programs of $\gamma$-ray selected blazars.

From the above results, we can derive an expected number of blazars which should be discovered by their high polarization and/or variation
in the fields of unidentified sources {\it if they are similar but unknown blazars}.
In total, 23 out of 26 blazars (88 \%) have pronounced polarization ($P > 10$ \%) or variation ($| H_{\rm IRSF} - H_{\rm 2MASS} | > 0.25$ mag).
Since 39 unidentified sources were observed, we would have $\sim$35 objects with such distinguishable properties under the above assumption.
However, $\sim$35 \% of them would be fainter than our limiting magnitude $H_{\rm IRSF}^{\rm lim} \simeq 16.0$ mag based on the 2MASS magnitude 
distribution of 1FGL blazars.
Furthermore, $\sim$32 \% of them would be outside SIRPOL field-of-view (7'.7 $\times$ 7'.7 arcmin$^2$) when $\gamma$-ray positions are used 
as the telescope pointing center, based on the distribution of distance between $\gamma$-ray positions and associated blazars.
Considering these restrictions, we expect to find 15 blazars out of 39 fields of unidentified sources if their $\gamma$-ray is indeed emitted 
by a similar population to 1FGL blazars.

\begin{figure}
\epsscale{1.0}
\plotone{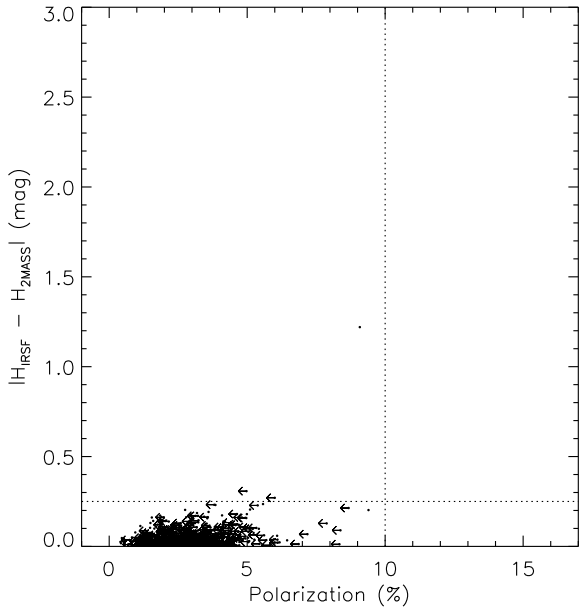}
\caption{Same as Figure \ref{var_stat_bzx}, but for the fields of unidentified sources. \label{var_stat_una}}
\end{figure}

Figure \ref{var_stat_una} shows polarization and variation of 608 objects detected in all the fields of unidentified sources.
While three objects are found to have high variation ($| H_{\rm IRSF} - H_{\rm 2MASS} | > 0.25$ mag), two of them with $P < 7$ \% and 
$| H_{\rm IRSF} - H_{\rm 2MASS} | \sim 0.3$ mag are likely contaminations; it is consistent with one contamination out of 326 objects 
in the blazar fields (Figure \ref{var_stat_bzx}).
On the other hand, the object at $P \sim 9$ \% and $| H_{\rm IRSF} - H_{\rm 2MASS} | \sim 1.2$ mag is a promising candidate of new $\gamma$-ray blazars.
It is found in J1315.6$-$0729 field, at R.A. 13$^{\rm h}$15$^{\rm m}$52$^{\rm s}$.98, Dec. $-$07$^{\rm d}$33$^{\rm m}$01$^{\rm s}$.99 
(J2000.0) with $H$-band brightness $H_{\rm IRSF} = 14.1$ mag and $H_{\rm 2MASS} = 15.3$ mag.
Its counterparts are found in the NED\footnote{
The NASA/IPAC Extragalactic Database (NED) is operated by the Jet Propulsion Laboratory, California Institute of Technology, under contract 
with the National Aeronautics and Space Administration.};
the optical variable source PQV1 J131553.00$-$073302.0 \citep{bauer09} and the radio source NVSS J131552$-$073301.
We plan to carry out a spectroscopic follow-up observation of this object in the near future.

Except for the above blazar candidate, we found no clear sign of variable or polarized objects in the fields of unidentified sources.
The apparent inconsistency between the expected number of blazar candidates as estimated above (15) and the actual number (1) indicates
that $\gamma$-ray emission of the unidentified sources arises from other types of objects than known 1FGL blazars.
They could be non-blazar active galaxies, starburst galaxies, or Galactic objects such as pulsars and supernova remnants at relatively
high Galactic latitudes, as well as active galaxies with jets but without strong polarization and variability at near-IR wavelengths.
In this regard, it is noteworthy that some of them are already identified (or associated) in the latest 2FGL catalog and the {\it Fermi}-LAT 
second AGN catalog \citep[2LAC;][]{ackermann11} as summarized in Table \ref{obslist_id_by_2fgl}.
Many of them are pulsars, which is consistent with our observation results.
While J2330.3$-$4745 is associated to the blazar PKS 2326$-$477, their separation is relatively large and the blazar is outside 
the field-of-view of our SIRPOL observation.

\begin{table*}
  \caption{2FGL identification (association) of 1FGL "unidentified" sources in our sample}\label{obslist_id_by_2fgl}
  \begin{center}
    \begin{tabular}{lll}
      \hline
      1FGL ID        & Associated source      & Object type  \\
      \hline
      J0001.9$-$4158 & 1RXS J000135.5$-$41551 & active galaxy of uncertain type\\
      J0101.0$-$6423 & PSR J0101$-$6422       & pulsar \\
      J0223.0$-$1118 & 1RXS J022314.6$-$11174 & active galaxy of uncertain type\\
      J0335.5$-$4501 & IRXS J033514.5$-$44592 & active galaxy of uncertain type\\
      J0614.1$-$3328 & PSR J0614$-$3330       & pulsar \\
      J1124.4$-$3654 & PSR J1124$-$36         & pulsar \\
      J1141.8$-$1403 & 1RXS J114142.2$-$14075 & active galaxy of uncertain type\\
      J1231.1$-$1410 & PSR J1231$-$1411       & pulsar \\
      J1312.6$+$0048 & PSR J1312$+$00         & pulsar \\
      J2241.9$-$5236 & PSR J2241$-$5236       & pulsar \\
      J2330.3$-$4745 & PKS 2326$-$477         & blazar \\
      \hline
    \end{tabular}
  \end{center}
\end{table*}



\acknowledgments

We are grateful to the IRSF team at Nagoya University, Kyoto University, and National Astronomical Observatory of Japan for great assistance 
provided during the observations.
This work was supported by Grant-in-Aid for Young Scientists (22684005) and the Global COE Program of Nagoya University 
"Quest for Fundamental Principles in the Universe" from JSPS and MEXT of Japan.


\begin{thebibliography}{}
\bibitem[Abdo et al.(2010a)]{2010ApJS..188..405A} Abdo, A.~A., Ackermann, 
M., Ajello, M., et al.\ 2010, \apjs, 188, 405 
 \bibitem[Abdo et al.(2010b)]{abdo10b} Abdo, A.~A., Ackermann, M., Ajello, M., et al.\ 2010b, \apj, 715, 429 
\bibitem[Ackermann et al.(2011a)]{ackermann11} Ackermann, M., Ajello, M., Allafort, A., et al.\ 2011a, \apj, 743, 171 
\bibitem[Ackermann et al.(2011b)]{ackermann11b} Ackermann, M., Ajello, M., Allafort, A., et al.\ 2011b, arXiv:1108.1202
\bibitem[Aharonian et al.(2006)]{aharonian06} Aharonian, F., Akhperjanian, A.~G., Bazer-Bachi, A.~R., et al.\ 2006, \nat, 440, 1018 
\bibitem[Bauer et al.(2009)]{bauer09} Bauer, A., Baltay, C., Coppi, P., et al.\ 2009, \apj, 705, 46 
 \bibitem[Bertin \& Arnouts(1996)]{1996A&AS..117..393B} Bertin, E., \& Arnouts, S.\ 1996, \aaps, 117, 393
 \bibitem[Heidt 
\& Nilsson(2011)]{2011A&A...529A.162H} Heidt, J., \& Nilsson, K.\ 2011, \aap, 529, A162 
 \bibitem[Kandori et al.(2006)]{2006SPIE.6269E.159K} Kandori, R., Kusakabe, N., Tamura, M., et al.\ 2006, \procspie, 6269,   
  \bibitem[Kovalev(2009)]{2009ApJ...707L..56K} Kovalev, Y.~Y.\ 2009, \apjl, 707, L56  
\bibitem[Matsumoto et al.(2005)]{matsumoto05} Matsumoto, T., Matsuura, S., Murakami, H., et al.\ 2005, \apj, 626, 31 
\bibitem[Matsuoka et al.(2011)]{matsuoka11a} Matsuoka, Y., Ienaka, N., Kawara, K., \& Oyabu, S.\ 2011, \apj, 736, 119 
\bibitem[Matsuoka et al.(2008)]{matsuoka08} Matsuoka, Y., Kawara, K., \& Oyabu, S.\ 2008, \apj, 673, 62 
\bibitem[Matsuoka et al.(2007)]{matsuoka07} Matsuoka, Y., Oyabu, S., Tsuzuki, Y., \& Kawara, K.\ 2007, \apj, 663, 781 
\bibitem[Matsuoka et al.(2012)]{matsuoka11b} Matsuoka, Y., Yuan, F.-T., Takeuchi, Y., \& Yanagisawa, K.\ 2012, \pasj, 64, 44
\bibitem[Nagashima et al.(1999)]{nagashima99} Nagashima, C., Nagayama, T., Nakajima, Y., et al.\ 1999, Star Formation 1999, 397 
\bibitem[Nagayama et al.(2003)]{nagayama03} Nagayama, T., Nagashima, C., Nakajima, Y., et al.\ 2003, \procspie, 4841, 459 
\bibitem[Nolan et al.(2012)]{nolan12} Nolan, P. L., et al., arXiv:1108.1435 
 \bibitem[Scarpa \& Falomo(1997)]{1997A&A...325..109S} Scarpa, R., \& Falomo, R.\ 1997, \aap, 325, 109 
 \bibitem[The Fermi-LAT Collaboration(2011)]{2011arXiv1108.1435T} The Fermi-LAT Collaboration 2011, arXiv:1108.1435 
\bibitem[Wardle \& Kronberg(1974)]{wardle74} Wardle, J.~F.~C., \& Kronberg, P.~P.\ 1974, \apj, 194, 249 
\end{thebibliography}
\end{document}